\documentclass[showkeys,showpacs]{revtex4}
\usepackage{amsmath}
\usepackage{amssymb}
\usepackage{graphicx}
\usepackage{color}

\begin{document}

\title{
A Star Harbouring a Wormhole at its Core
}

\author{
Vladimir Dzhunushaliev,$^{1,2,3}$
\footnote{
Email: vdzhunus@krsu.edu.kg}
Vladimir Folomeev,$^{2,3}$
\footnote{Email: vfolomeev@mail.ru}
Burkhard Kleihaus,$^{3}$
\footnote{Email: kleihaus@theorie.physik.uni-oldenburg.de }
Jutta Kunz$^3$
\footnote{Email:  kunz@theorie.physik.uni-oldenburg.de}
}
\affiliation{$^1$Institute for Basic Research,
Eurasian National University,
Astana, 010008, Kazakhstan
 \\ 
$^2$Institute of Physicotechnical Problems and Material Science of the NAS
of the
Kyrgyz Republic, 265 a, Chui Street, Bishkek, 720071,  Kyrgyz Republic \\
$^3$Institut f\"ur Physik, Universit\"at Oldenburg, Postfach 2503
D-26111 Oldenburg, Germany
}

\begin{abstract}
We consider a configuration consisting of a wormhole
filled by a perfect fluid.
Such a model can be applied to describe stars as well as
neutron stars with a nontrivial topology.
The presence of a tunnel allows for motion of the fluid,
including oscillations near the core of the system.
Choosing the polytropic equation of state for the perfect fluid,
we obtain static regular solutions.
Based on these solutions, we consider small radial oscillations
of the configuration
and show that the solutions are stable with respect to linear perturbations
in the external region.
\end{abstract}

\maketitle

\section{Introduction}

After the discovery of the accelerated expansion
of the present Universe at the end of the 1990ies,
a great deal of attention has focused on models
in which violation of one of the energy conditions takes place.
In hydrodynamical language, this means that the
parameter $w$ in the equation of state $p=w \rho$,
defining the ratio of the pressure $p$ of some matter
to its energy density, may
be less than $-1/3$ (i.e.~violation of the strong energy condition),
or even less than $-1$
(i.e.~violation of the weak energy or null energy condition).
While $w=-1$ is realized by the presence of a cosmological constant,
leading to exponential expansion of the Universe,
$w<-1$ would be realized by phantom matter with a possible
end of the Universe in a finite time.
Recent astronomical observations indicate that such a
possibility is not excluded
\cite{Tonry:2003zg,Alam:2003fg}.

On the other hand, when considering models of compact astrophysical objects,
a violation of the weak/null energy condition
leads to a possible existence of configurations with non-trivial topology
 -- traversable wormholes \cite{Thorne:1988}.
At the present time, much work has been devoted
to the study of various models of both microscopic and
macroscopic wormholes.
One of the possible variants of obtaining wormhole-like solutions
consists in considering ghost scalar fields,
i.e.~scalar fields with the
opposite sign in front of the kinetic energy term.
The use of such fields allows, in some cases, to obtain an
equation of state with $w<-1$, and thus to violate
the weak/null energy condition.

Models with ghost scalar fields have been considered earlier.
The early pioneering works with a massless ghost scalar field have been done in Refs. \cite{Bronnikov:1973fh,Ellis:1973yv}.
A ghost scalar field with a Mexican hat potential was investigated in the papers \cite{Kodama:1978dw,Kodama:1979}, where
it was found that this system had regular, stable solutions only
for topologically non-trivial (wormhole-like) geometry.
In
\cite{Kuhfittig:2010wz,lxli,ArmendarizPicon:2002km,Sushkov:2002ef,Lobo:2005us,Sushkov:2005kj,Smith:1998qx}
traversable Lorentzian wormholes were investigated further,
refining the conditions on the type of matter or fields
that would lead to such space-times.
A general overview of the subject of Lorentzian wormholes and
violations of the various energy conditions
can be found in the book by Visser \cite{Visser}.

In the case of macroscopic wormholes, ghost fields can be used
to create models of astrophysical objects
which can serve as entrances to wormholes.
 From the point of view of a distant observer,
such objects will be quite similar to the usual
star-like configurations,
but they will have some characteristic features.
Moreover, in the presence of electric and/or magnetic charges
there exist new possibilities to detect such objects via
galactic and extragalactic observations.
A discussion on the observability of such
electrically/magnetically charged wormholes
is given in \cite{Kardashev:2006nj}.

However, it is also possible to imagine a situation
where the concentration of exotic matter
violating the weak/null energy condition
is formed at the center of an otherwise ordinary star.
This in turn then leads to the possibility of the creation of a tunnel --
a wormhole whose throat is filled with ordinary (star) matter.
Thus an ordinary star or a neutron star could
harbour some exotic matter at its core
providing a nontrivial wormhole topology.
Such a combined configuration with a nontrivial spacetime topology differs
from the usual spherically symmetric stars consisting of ordinary (nonexotic)
matter and having a center at the point $r=0$. In the model considered here,
there is no such center but there exists some minimal value of the radial
coordinate $r=r_{min}$ corresponding to the radius of the throat.
But such a configuration
will still be different from a usual wormhole due to the presence of ordinary matter.
Taking this into account, we will describe the region around the throat by
using the term ``core'' along with the term ``throat''.
The advantage of introducing
this term is that on the one hand it
implies the presence of the throat and, on the other hand, it allows to keep
the spirit of considering star-like configurations consisting only of ordinary matter.

For a distant observer, such a configuration would very much look
like an ordinary star.
However, also some distinctions should be present. For example:

(i) One of the most striking differences is that instead of a single star
two similar stars will be observed that are
separated in space.
These two stars are associated with the two mouths of the wormhole.

(ii) Due to the fact that matter inside such star can move freely
through the tunnel, the presence of radial
(including quasiperiodical) motion is possible.
Thus one can estimate the energy associated with such motion.

In this connection, we here suggest to consider a simple (toy) model
for a ``wormhole plus ordinary matter'',
where the source of the exotic matter is a massless scalar field,
and the ordinary matter is a perfect fluid
with a polytropic equation of state $p=k\rho^\gamma$.
In section~\ref{wh_met} we consider the general
properties of this model.
In section~\ref{stat_sol} we find static solutions
for the perfect polytropic fluid filling the wormhole's throat.
In section~\ref{rad_oscill} the infinitesimal radial oscillations
of the system are considered.
In Appendix \ref{appen_transf} the equations
for the perfect polytropic fluid are derived following standard notation.
Finally, in Appendix \ref{appen_osc_poly} the oscillating solutions
for the perfect fluid with a polytropic equation of state are found;
while in Appendix C stability of the external solution
under linear perturbations is addressed.

\section{General properties of the model
}
\label{wh_met}
Let us first consider some general properties of the
model consisting of a gravitating massless ghost scalar field
$\phi$ with negative kinetic term
and a normal isotropic perfect fluid.
The Lagrangian of the system is chosen in the form
\begin{equation}
\label{lagran_wh_star}
L=-\frac{R}{16\pi G}-\frac{1}{2}\partial_{\mu}\phi\partial^{\mu}\phi +L_m,
\end{equation}
where $L_m$ refers to the isotropic perfect fluid.
The energy-momentum tensor of this system is then
\begin{equation}
\label{emt_wh_star}
T_i^k=(\rho+p)u_i u^k-\delta_i^k p-\partial_{i}\phi\partial^{k}\phi+\frac{1}{2}\delta_i^k
\partial_{\mu}\phi\partial^{\mu}\phi,
\end{equation}
where $\rho$ and $p$ are the energy density and the pressure of the fluid,
$u^i$ is the four-velocity (in units where $c=1$).

For the metric of this system we employ Schwarzschild-like coordinates
\begin{equation}
\label{metric_wh}
ds^2=e^{\nu(r)}dt^2-\frac{dr^2}{1-b(r)/r}-r^2\left(d\theta^2+\sin^2\theta d\phi^2\right).
\end{equation}
The general properties of such a metric describing a wormhole can be found
in Refs.~\cite{Thorne:1988,Sushkov:2005kj,Visser}:

(i) the coordinate $r$ covers the range $r_0 \leq r < +\infty$, where $r_0$ is the throat radius;

(ii) in order to cover the whole space-time one needs to use two copies of the coordinate system \eqref{metric_wh};

(iii) the requirement of the absence of horizons or singularities leads to the condition that $\nu$ should be everywhere finite;

(iv) throughout space-time $(1-b(r)/r)\geq 0$ implying
$$b(r_0)=r_0, \quad b^\prime(r_0)<1, \quad b(r)<r;$$

(v) to provide asymptotic flatness of the system,
it is necessary to satisfy the asymptotic limit
$$ b(r)/r \to 0 \quad \text{as} \quad |r|\to\infty.$$

Taking into account these general properties
of the model under consideration,
in studying the ``wormhole plus star'' system
it is necessary, first of all,
to choose some equation of state for the fluid filling the wormhole.
As an example, we next consider a perfect fluid
with a polytropic equation of state,
which is of a considerable interest in astrophysical applications.

\boldmath
\section{Perfect fluid with the equation of state $p=k\,\rho^\gamma$}
\unboldmath
\label{stat_sol}

We now consider the case
that the fluid filling the wormhole has a polytropic equation of state.
In our calculations, we will follow Ref.~\cite{Tooper:1964},
where the model of a general relativistic polytropic fluid sphere
is considered.
The equation of state for the polytropic fluid is
\begin{equation}
\label{eqs_fluid}
p=k\,\rho^\gamma,
\end{equation}
where $k, \gamma$ are constants,
and $\rho$ is the mass density of the fluid.
(The mass density corresponds to the energy density because of $c=1$.)
Follow Ref.~\cite{Tooper:1964},
let us rewrite this equation in terms of the variable
$\theta$ introduced as
\begin{equation}
\label{rho_theta}
\rho=\rho_c \theta^n,
\end{equation}
where $\rho_c$ is the core density,
and the constant $n$, called the polytropic index,
is related to $\gamma$ via $n=1/(\gamma-1)$.
Then Eq.~\eqref{eqs_fluid} takes the form
\begin{equation}
\label{eqs_fluid_theta}
p=k\,\rho^\gamma=k \rho^{1+1/n}=k\rho_c^{1+1/n} \theta^{n+1}.
\end{equation}

Using this equation of state,
one finds the following system of equations
(for details, see Appendix \ref{appen_transf}):
\begin{eqnarray}
\label{eq_theta}
	\xi^2\frac{1-\frac{2\sigma(n+1)v}{\xi}}{1+\sigma\theta}\frac{d\theta}{d\xi} &=&
	\xi^3\left[\theta^n\left(1-\sigma\theta\right)-
	\frac{1}{\xi^2}\frac{d v}{d\xi}\right]-v
,\\
\label{eq_v}
	\frac{d v}{d\xi} &=& \xi^2\left[
	\theta^n -	\frac{1}{2}\frac{\bar{D}^2}{\xi^4}e^{-\nu_c}
	\left(
		\frac{1+\sigma}{1+\sigma\theta}
	\right)^{-2(n+1)}
	\right],
\end{eqnarray}
where $\bar{D}$ is the dimensionless constant
$$
\bar{D}=\frac{4\pi G D}{(n+1)\sigma} \sqrt{\rho_c},
$$
$\xi$ is the dimensionless radius
\begin{equation}
\label{rad_dless}
\xi=A r, \quad A=\left[\frac{4\pi G \rho_c}{(n+1)k \rho_c^{1/n}}\right]^{1/2},
\end{equation}
and $v(\xi)$ is the dimensionless function
\begin{equation}
\label{v_dless}
v(\xi)=\frac{A^3 M}{4\pi \rho_c} u(r).
\end{equation}
The function $u(r)$ is a new metric function determining the component
$g_{rr}$ [see Appendix \ref{appen_transf}, Eq. \eqref{u_app}],
and $M$ is the total mass of the configuration within the range
$\xi_0 \leq \xi \leq \xi_b$, where $\xi_0$ is the throat radius,
and $\xi_b$ is the boundary of the fluid where $\theta=0$.
The constant $\sigma$ is defined by the formula \eqref{sigma_expr},
and is equal to the ratio of the core pressure $p_c$
and the core density $\rho_c$ of the fluid.

Equations \eqref{eq_theta} and \eqref{eq_v}
are to be solved for given $n$ and $\sigma$
subject to the boundary conditions
\begin{equation}
\label{bound}
\theta_0\equiv\theta(\xi_0)=1, \quad v_0\equiv v(\xi_0)=\frac{1}{2(n+1)\sigma}\xi_0.
\end{equation}
The first condition corresponds to the fact
that the core density is $\rho_c$,
and the second condition follows from the fact
that at the throat $b(r_0)=r_0$
(see condition (iv) of the previous section).
Then, by using the definition of $b(r)$
through $u(r)$ (see Eq.~\eqref{u_app})
\begin{equation}
\label{metr_b}
b(r)=2 G M u(r),
\end{equation}
one obtains condition \eqref{bound}.

The boundary condition for $v(\xi_0)$ implies
that the coefficient $(1-2\sigma(n+1)v/\xi)$
in front of the derivative of the function $\theta$
in Eq.~\eqref{eq_theta} goes to zero
at the throat,
which, in general, leads to the occurrence of a singularity in the solution.
To avoid this, expand the solution of the system
\eqref{eq_theta}-\eqref{eq_v}
in the neighborhood of the point
$\xi=\xi_0$ and start calculating the solution at $\xi=\xi_1$,
choosing the boundary condition for $v$ in the form
\begin{equation}
\label{bound_v}
v(\xi_1)=v_0+v_1(\xi_1-\xi_0).
\end{equation}
Substituting this into Eq. \eqref{eq_v},
we find the following expression for $v_1$:
\begin{equation}
\label{const_v1}
v_1= \xi_0^2\left[\theta_0^n-\frac{1}{2}\frac{\bar{D}^2}{\xi_0^4}e^{-\nu_c}\left(\frac{1+\sigma}{1+\sigma\theta_0}\right)^{-2(n+1)}\right].
\end{equation}
Further, to exclude the appearance of a singularity in Eq.~\eqref{eq_theta},
it is necessary to assume that, together with $(1-2\sigma(n+1)v/\xi)$,
the right hand side of Eq.~\eqref{eq_theta} simultaneously goes to zero.
Proceeding from this requirement,
we obtain the following expression for $\bar{D}$:
$$
\bar{D}^2=2\xi_0^4 e^{\nu_c}\left(\frac{1+\sigma}{1+\sigma\theta_0}\right)^{2(n+1)}
\left\{\theta_0^n-\frac{1}{\xi_0^3}\left[\xi_0^3\theta_0^n(1-\sigma\theta_0)-v_0\right]\right\}.
$$
Substituting this into Eq.~\eqref{const_v1}, we find
\begin{equation}
\label{const_v1_f}
v_1= \frac{1}{\xi_0}\left[\xi_0^3 \theta_0^n(1-\sigma\theta_0)-v_0\right].
\end{equation}
Thus, the mathematical description of an equilibrium static configuration
is achieved by means of the system of equations \eqref{eq_theta}-\eqref{eq_v}
with the boundary conditions Eq.~\eqref{bound} for $\theta_0$,
and Eqs.~\eqref{bound_v} and \eqref{const_v1_f} for $v(\xi_1)$.

\begin{figure}[t]
\centering
  \includegraphics[height=6cm]{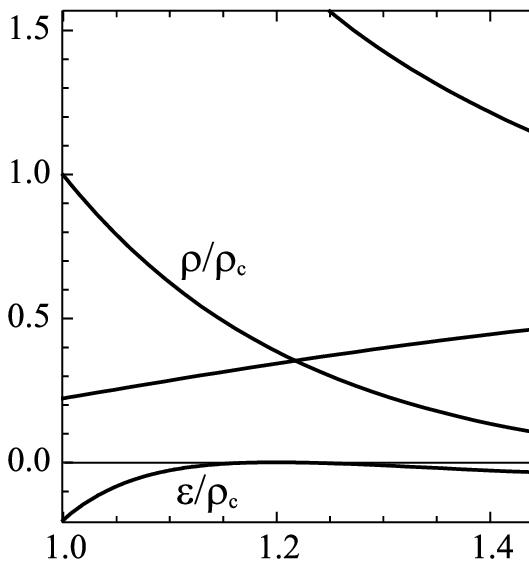}
\caption{The metric functions $g_{tt}=e^{\nu}$, $\log{|g_{rr}|}=\lambda$,
the fluid density $\rho$, and the total energy density $\varepsilon$
from \eqref{tot_dens} for the polytropic index $n=1.5$.
The point where $\rho=0$ corresponding to the boundary of the fluid
is situated at $\xi_b \approx 1.79$.
To provide the asymptotical flatness of the solutions,
i.e.~$e^\nu, e^\lambda \to 1$ as $\xi \to \infty$,
the value of the constant $\nu_c$ should be chosen as
$\nu_c \approx -1.46$.}
\label{n_1_5}
\end{figure}

\begin{figure}[t]
\centering
  \includegraphics[height=6cm]{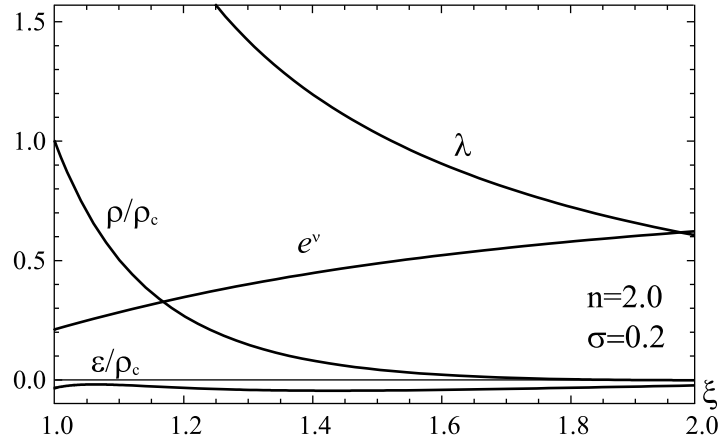}
\caption{The metric functions $g_{tt}=e^{\nu}$, $\log{|g_{rr}|}=\lambda$,
the fluid density $\rho$, and the total energy density $\varepsilon$
from \eqref{tot_dens} for the polytropic index $n=2.0$.
The point where $\rho=0$ corresponding to the boundary of the fluid
is situated  at $\xi_b \approx 1.99$.
To provide the asymptotical flatness of the solutions,
i.e.~$e^\nu, e^\lambda \to 1$ as $\xi \to \infty$,
the value of the constant $\nu_c$ should be chosen as
$\nu_c \approx -1.57$.}
\label{n_2_0}
\end{figure}

Examples of numerical solutions obtained in this way are presented
in Figs.~\ref{n_1_5} and \ref{n_2_0}
for the polytropic index $n=1.5$ and $n=2$, respectively.
In these figures, the following functions are shown:
the metric functions $g_{tt}=e^{\nu}$ and $g_{rr}=-e^\lambda$;
the distribution of the energy density of the fluid $\rho$,
and the total energy density $\varepsilon \equiv T_0^0$
of the system consisting of the scalar field and the fluid.
(Both $\rho$ and $\varepsilon$ are given
in units of the core density $\rho_c$.)

For the graphs, we have employed the following expressions:
$e^{\nu}$ is obtained from Eq.~\eqref{nu_app}
\begin{equation}
\label{nu_poly}
e^{\nu}=e^{\nu_c}\left[\frac{1+\sigma}{1+\sigma \theta}\right]^{2(n+1)}.
\end{equation}
$e^\lambda$ follows from Eq. \eqref{u_app},
using \eqref{rad_dless} and \eqref{v_dless}
\begin{equation}
\label{lambda_poly}
e^\lambda=\left[1-2\sigma(n+1)\frac{v}{\xi}\right]^{-1}
\end{equation}
(note that this expression diverges at $\xi=\xi_0$,
as discussed in connection with the boundary conditions),
and $\varepsilon$ follows
by using \eqref{rho_theta}, \eqref{rad_dless} and \eqref{nu_poly}
\begin{equation}
\label{tot_dens}
\varepsilon \equiv T_0^0=\rho-\frac{1}{2} \frac{D^2}{r^4}e^{-\nu}=
\rho_c\left[\theta^n-\frac{1}{2}\frac{\bar{D}^2 e^{-\nu_c}}{\xi^4}\left(\frac{1+\sigma}{1+\sigma \theta}\right)^{-2(n+1)}\right].
\end{equation}

The integration constant $\nu_c$, corresponding
to the value of $\nu$ at  the throat,
is determined by requiring $e^{\nu}$ to be equal to unity at infinity,
corresponding to an asymptotically flat space-time.
To fix the constant, it is necessary to find the external solution
of the field equations outside the fluid.
For this purpose, let us use the Einstein equations
\eqref{Einstein-00_app_n} and \eqref{Einstein-11_app_n},
and the first integral \eqref{first_int},
taking into account that in this region there is no ordinary matter,
i.e.~$\theta=0$.
This leads to the following system of equations:
\begin{eqnarray}
\label{Einstein-00_ext}
&&-e^{-\lambda}\left(\frac{1}{r^2}-\frac{\lambda^\prime}{r}\right)+\frac{1}{r^2}
=-4\pi G\frac{D^2}{r^4} e^{-\nu},
 \\
\label{Einstein-11_ext}
&&-e^{-\lambda}\left(\frac{1}{r^2}+\frac{\nu^\prime}{r}\right)+\frac{1}{r^2}
=4\pi G\frac{D^2}{r^4} e^{-\nu},\\
\label{first_int_ext}
&&\phi^{\prime 2}=\frac{D^2}{r^4}e^{\lambda-\nu},
\end{eqnarray}
which, by analogy with the transformations made above,
can be rewritten in terms of the dimensionless variables
$v(\xi), \varphi(\xi)$ and $\nu(\xi)$ as follows
\begin{eqnarray}
\label{eq_v_ext}
	\frac{d v}{d\xi} &=& -\frac{1}{2}\frac{\bar{D}^2}{\xi^2}e^{-\nu},
 \\
\label{eq_nu_ext}
	\frac{d\nu}{d\xi} &=&
	\frac{1}{\xi}\left[
		\frac{1-\sigma (n+1)\frac{\bar{D}^2}{\xi^2}e^{-\nu}}{1-2\sigma (n+1) \frac{v}{\xi}}
	-1\right],
\\
\label{eq_phi_ext}
	\left(\frac{d\varphi}{d\xi}\right)^2 &=&
	\frac{\bar{D}^2}{\xi^4}
	\frac{e^{-\nu}}{1-2\sigma (n+1) \frac{v}{\xi}}.
\end{eqnarray}
This system contains the parameter $\sigma$ as a trace
of the influence of the fluid on the external solution.
The solution is to be sought beginning from the surface of the star
at $\xi=\xi_b$ by using, as the boundary conditions,
the values of $v(\xi_b), \varphi(\xi_b)$ and $\nu(\xi_b)$
obtained from the solution of the equations
\eqref{eq_theta} and \eqref{eq_v}
for the internal part of the configuration.
This then allows to determine the value of the integration constant $\nu_c$
by requiring $e^{\nu}$ to be equal to unity at infinity,
providing asymptotical flatness of the space-time.
(The values of $\nu_c$ for the examples shown in
figures \ref{n_1_5} and \ref{n_2_0} are given in the captions.)

The systems of equations \eqref{eq_theta}-\eqref{eq_v}
and \eqref{eq_v_ext}-\eqref{eq_phi_ext}
can be used to determine the total mass of the system under consideration.
For the spherically symmetric metric \eqref{metric_wh} the function
$b(r)$ is associated with the effective mass $m(r)$ inside the radius $r$
as follows \cite{Thorne:1988,Sushkov:2005kj,Visser}
$$
m(r)=b(r)/2=r_0/2+4\pi\int_{r_0}^{r} \varepsilon(x)x^2 dx.
$$
Here the integration constant is chosen to provide $b(r_0)=r_0$
(see condition (iv) from section \ref{wh_met}),
and $\varepsilon(x)$ is taken from Eq.~\eqref{tot_dens}.
Using formulas \eqref{v_dless} and \eqref{metr_b},
we introduce the dimensionless function
$$
B(\xi)\equiv A\, b=2\sigma (n+1) v(\xi)
$$
defining the dimensionless mass function of the configuration as
$M(\xi)=B(\xi)/2$.
Then the asymptotic value $\lim_{\xi\to \infty}M(\xi)=M_\infty$
corresponds to the total mass of the configuration.
For the cases presented in figures \ref{n_1_5} and \ref{n_2_0},
the total masses are $M_{n=1.5}\approx M_{n=2.0}\approx 0.37$.
Thus despite the violation of the energy conditions
and the presence of a negative energy density,
the total mass of the configurations remains positive.

\section{Radial oscillations of the system}
\label{rad_oscill}

We now study the radial oscillations of the above static system.
Bearing in mind the possibility
of using an arbitrary equation of state $p=p(\rho)$,
we will derive perturbation equations
without imposing any preliminary restrictions on the equation of state.
Various general-relativistic stellar models
have been thoroughly investigated.
In particular, infinitesimal radial oscillations of a gas sphere
with polytropic equation of state has been discussed
by Chandrasekhar \cite{Chandrasekhar:1964zz}.
Here we employ the techniques developed in \cite{Chandrasekhar:1964zz}
for investigating the infinitesimal radial oscillations
of the fluid filling the wormhole.

To this end,
it is more convenient to choose the metric in the form
(following Ref.~\cite{Chandrasekhar:1964zz}):
\begin{equation}
\label{metric_oscil}
ds^2=e^{\nu}dt^2-e^{\lambda} dr^2-r^2\left(d\theta^2+\sin^2\theta d\phi^2\right),
\end{equation}
where the metric functions $\nu, \lambda$ depend both on time $t$
and radial coordinate $r$.
The corresponding Einstein equations can be written in the form
\begin{eqnarray}
\label{Einstein-00osc}
&&
-e^{-\lambda}\left[\frac{1}{r^2}-\frac{\lambda^\prime}{r}\right]+\frac{1}{r^2}=
-\frac{1}{r^2}\frac{\partial}{\partial r}\left[r e^{-\lambda}\right]+\frac{1}{r^2}=8\pi GT^0_0,\\
\label{Einstein-11osc}
&&
-e^{-\lambda}\left[\frac{1}{r^2}+\frac{\nu^\prime}{r}\right]+\frac{1}{r^2}=8\pi G T_1^1,\\
\label{Einstein-10osc}
&&-e^{-\lambda}\frac{\dot{\lambda}}{r}=8\pi G T_0^1.
\end{eqnarray}
In the above equations, ``prime" and ``dot" denote differentiation
with respect to  $r$ and $t$, respectively.
Considering only infinitesimal oscillations
the components of the four-velocity are given by
$$
u^0=e^{-\nu_0/2}, \quad u_0=e^{\nu_0/2}, \quad u^1=e^{-\nu_0/2} v, \quad u_1=-e^{\lambda_0-\nu_0/2} v,
$$
with three-velocity
$$
v=\frac{d r}{d t} \ll 1,
$$
and the index 0 of the metric functions indicates
the static background solutions of the Einstein equations.
The components of the energy-momentum tensor \eqref{emt_wh_star} are then
given by
\begin{eqnarray}
\label{emt-00}
&&
T_0^0=\rho-\frac{1}{2} e^{-\nu}\dot{\phi}^2-\frac{1}{2} e^{-\lambda}\phi^{\prime 2}
,\\
\label{emt-11}
&&
T_1^1=-p+\frac{1}{2} e^{-\nu}\dot{\phi}^2+\frac{1}{2} e^{-\lambda}\phi^{\prime 2},\\
\label{emt-10}
&&
T_0^1=(\rho_0+p_0)v+ e^{-\lambda}\dot{\phi}\,\phi^\prime.
\end{eqnarray}

Let us seek perturbed solutions of the form
\begin{equation}
\label{perturbations}
y=y_0+y_p,
\end{equation}
where the index 0 refers to static solutions,
the index $p$ indicates the perturbation,
and $y$ denotes one of the functions $\lambda, \nu, \rho, p$ or $\phi$.
Substituting these expressions into
Eqs.~\eqref{Einstein-00osc} and \eqref{Einstein-11osc}, we find:
\begin{eqnarray}
\label{Einstein-00pert}
&&
e^{-\lambda_0}\left[r \lambda_p^\prime+\lambda_p\left(1-r \lambda_0^\prime\right)\right]\equiv
\frac{\partial}{\partial r}\left[r e^{-\lambda_0} \lambda_p\right]=
8\pi G r^2\left[\rho_p-e^{-\lambda_0}\phi_0^\prime\left(\phi_p^\prime-\frac{1}{2}\phi_0^\prime \lambda_p\right)\right]
,\\
\label{Einstein-11pert}
&&
e^{-\lambda_0}\left[\frac{\nu_p^\prime}{r}-\frac{\nu_0^\prime}{r}\lambda_p-\frac{\lambda_p}{r^2}\right]=
8\pi G\left[-p_p+e^{-\lambda_0}\phi_0^\prime\left(\phi_p^\prime-\frac{1}{2}\phi_0^\prime \lambda_p\right)\right].
\end{eqnarray}
Next, from Eq.~\eqref{Einstein-10osc} we have
\begin{equation}
\label{Einstein-10pert}
e^{-\lambda_0}\frac{\dot{\lambda}_p}{r}=-8\pi G\left[(\rho_0+p_0)v+ e^{-\lambda_0}\dot{\phi_p}\,\phi_0^\prime\right].
\end{equation}
The Einstein field equations
are not all independent since 
$$
T^k_{i;k}=0.
$$
The $i=1$ component of this equation has the form
\begin{equation}
\label{conserv_osc}
\frac{\partial T^0_1}{\partial t}+\frac{\partial T^1_1}{\partial r}+\frac{1}{2}\left(\dot{\nu}+\dot{\lambda}\right)T^0_1+
\frac{1}{2}\left(T_1^1-T_0^0\right)\nu^\prime+\frac{2}{r}\left[T_1^1-\frac{1}{2}\left(T^2_2+T^3_3\right)\right]=0.
\end{equation}
In the model under consideration
the perturbed components $T^2_2$ and $T^3_3$ are given by
\begin{equation}
\label{emt-22}
T^2_2=T_3^3=-p_p-e^{-\lambda_0}\phi_0^\prime\left(\phi_p^\prime-\frac{1}{2}\phi_0^\prime \lambda_p\right).
\end{equation}
Now it is convenient to introduce a ``Lagrangian displacement'' $\zeta$
with respect to time $t$ defined by
$$
v=\frac{\partial \zeta}{\partial t}.
$$
Then Eq.~\eqref{Einstein-10pert} can directly be integrated to give
\begin{equation}
\label{lambda_pert}
\lambda_p=-8\pi G e^{\lambda_0} r \left[\left(\rho_0+p_0\right)\zeta+e^{-\lambda_0}\phi_0^\prime \phi_p\right].
\end{equation}
Using this expression, we have from  Eq.~\eqref{Einstein-00pert}
\begin{equation}
\label{E_pert}
\rho_p=e^{-\lambda_0}\phi_0^\prime \left(\phi_p^\prime-\frac{1}{2}\phi_0^\prime \lambda_p\right)-
\frac{1}{r^2}\frac{\partial}{\partial r}\Big\{r^2\left[\left(\rho_0+p_0\right)\zeta+e^{-\lambda_0}\phi_0^\prime \phi_p\right]\Big\},
\end{equation}
and the corresponding expression for $\nu_p^\prime$ follows
from Eq.~\eqref{Einstein-11pert}
\begin{equation}
\label{nu_prime_pert}
\nu_p^\prime=\nu_0^\prime \lambda_p+\frac{\lambda_p}{r}-8\pi G r e^{\lambda_0}
\left[-p_p+e^{-\lambda_0}\phi_0^\prime\left(\phi_p^\prime-\frac{1}{2}\phi_0^\prime \lambda_p\right)\right].
\end{equation}

Now we suppose that all perturbations
have the following harmonic dependence on time $t$
\begin{equation}
\label{harmonic}
y_p(t,r) = \bar{y}_p(r) e^{i\omega t},
\end{equation}
where the function
$\bar{y}_p(r)$ depends only on the space coordinate $r$, and
the characteristic frequency $\omega$ should be determined
from the calculations.
(For convenience, we hereafter drop the bar.)
Then, using the expressions \eqref{emt-00}-\eqref{emt-10} and \eqref{emt-22},
the perturbed equation \eqref{conserv_osc} takes the form
\begin{align}
\label{conserv_osc_pert}
\begin{split}
&\omega^2 e^{\lambda_0-\nu_0}\left[\left(\rho_0+p_0\right)\zeta+e^{-\lambda_0}\phi_0^\prime \phi_p\right]-
\frac{\partial p_p}{\partial r}+e^{-\lambda_0}\Big\{\left(\phi_0^{\prime\prime}-\lambda_0^\prime \phi_0^\prime\right)\left(
\phi_p^\prime-\frac{1}{2}\phi_0^\prime \lambda_p\right)+\phi_0^\prime\left[\phi_p^{\prime\prime}-
\frac{1}{2}\left(\phi_0^{\prime\prime}\lambda_p+\phi_0^{\prime}\lambda_p^\prime\right)\right]\Big\}\\
&-\frac{1}{2}\left(\rho_p+p_p\right)\nu_0^\prime+e^{-\lambda_0}\left[\phi_0^\prime \nu_0^\prime \phi_p^\prime+
\frac{1}{2}\phi_0^{\prime 2}\left(\nu_p^\prime-\nu_0^\prime \lambda_p\right)\right]+
\frac{4}{r}e^{-\lambda_0}\phi_0^\prime \left(\phi_p^\prime-\frac{1}{2}\phi_0^\prime \lambda_p\right)-
\frac{1}{2}\left(\rho_0+p_0\right)\nu_p^\prime=0.
\end{split}
\end{align}

By adding to this equation the equation for the perturbed scalar field
$\phi_p$ which follows from
$$
\frac{1}{\sqrt{-g}}\frac{\partial}{\partial x^i}\left[\sqrt{-g}g^{ik}\frac{\partial \phi}{\partial x^k}\right]=0,
$$
and takes the form
\begin{equation}
\label{phi_pert}
\phi_p^{\prime\prime}+\left[\frac{2}{r}+\frac{1}{2}\left(\nu_0^\prime-\lambda_0^\prime\right)\right]\phi_p^\prime+
\frac{1}{2}\left(\nu_p^\prime-\lambda_p^\prime\right)\phi_0^\prime+\omega^2 e^{\lambda_0-\nu_0}\phi_p=0,
\end{equation}
we have a system of two second-order ordinary differential equations
\eqref{conserv_osc_pert} and \eqref{phi_pert}
with respect to the scalar field perturbation $\phi_p$
and displacement $\zeta$
which describes the infinitesimal radial oscillations of the system
``wormhole plus star''.

Thus, substituting equations
\eqref{lambda_pert}-\eqref{nu_prime_pert}, \eqref{press_pert}
and \eqref{press_der_pert}
to the system \eqref{conserv_osc_pert} and \eqref{phi_pert},
{\it regular} solutions of the indicated system are to be sought.
As shown in previous investigations of such a type of problem,
the solution of an equation similar to Eq.~\eqref{conserv_osc_pert},
even without a scalar field, is a non-trivial problem \cite{Bardeen:1966}.
In the latter case, when considering
an ordinary spherically symmetric star-like configuration (where $r_0=0$),
the boundary conditions are chosen in the following form
\cite{Chandrasekhar:1964zz}:
$$
\zeta=0 \quad \text{at} \quad r=0 \quad \text {and} \quad p_p=0 \quad \text{at} \quad r=R,
$$
where $R$ is the radius of a star.
As pointed out in \cite{Bardeen:1966},
in the case of a polytropic equation of state
which is widely used in the construction of hot ($T>0$) stellar models
of physical interest, $\gamma$ is finite and non-zero at $r=R$,
and $\rho_0/p_0$ has a pole of order 1.
In such a situation the point $r=R$ is a regular singular point
of the eigenequation \eqref{conserv_osc_pert},
and one should choose the value of the derivative $(d\zeta/dr)_{r=R}$
to provide regularity of the solution at $r=R$.
The further solution then reduces to employing one of the methods
described in Ref.~\cite{Bardeen:1966}.

The numerical calculations for the case of a polytropic equation of state
presented in Appendix \ref{appen_osc_poly} show
that in the presence of a wormhole
the situation on the boundary of the fluid
remains similar to the case of an ordinary star:
at $r=R$ there is the divergence of $\zeta$.

\section{Conclusions and remarks}

The main purpose of the paper has been the demonstration
that regular static solutions exist
for the system of a star with a wormhole,
describing self-consistently a configuration composed of
a gravitating perfect fluid and exotic matter at its core.
The presence of the exotic matter allows for the existence
of the non-trivial topology.
As exotic matter, a massless ghost scalar field has been chosen
(see the Lagrangian \eqref{lagran_wh_star}).
The perfect fluid has been modeled
by a polytropic equation of state \eqref{eqs_fluid}.
For this system, by means of numerical calculations,
we have obtained regular solutions, which we demonstrated for
the parameters $\gamma=5/3$ and $\gamma=1.5$
(see Figs.~\ref{n_1_5} and \ref{n_2_0}, respectively).

Our general procedure to obtain the solutions is summarized as follows:
we start from the core of the configuration
(from the throat of the wormhole)
with some initial core density of the fluid
corresponding to the density at the throat.
This core density can be chosen either as the density
of an ordinary non-relativistic star
or, for instance, as the density of a neutron star.
The condition of regularity of the solutions at the core
then requires an appropriate choice of the boundary conditions
(see Eqs.~\eqref{bound}-\eqref{const_v1_f}).
The subsequent behavior of the solutions is determined by the
model's parameters:
the polytropic index $n$ and the parameter $\sigma$
which  denotes the ratio of the core pressure $p_c$
and the core density $\rho_c$ of the fluid.
As the dimensionless radial coordinate $\xi$ increases,
the density of the matter $\theta$ decreases,
reaching zero at some boundary value $\xi=\xi_b$.
This value of the radial coordinate can be considered as the boundary of
the ordinary matter.
 From the point of view of a distant observer,
this is a visible boundary of the star.
Note, that in this model
the scalar field $\phi$ and its derivative $\phi^\prime$
are not yet equal to zero on the boundary of the fluid.
It requires matching of the internal fluid solutions with external solutions
having a nonzero scalar field energy density,
that vanishes asymptotically. 

The essential point is that this model differs from ordinary stars
by the presence of a tunnel at the star's core.
It is obvious that, in principle,
the matter can move freely through the tunnel.
To investigate such motion, one can use the approach suggested in
Ref.~\cite{Chandrasekhar:1964zz}
to describe infinitesimal radial oscillations of the gas
within the framework of general relativity.
Applying this method to the ``star plus wormhole" system,
we have derived the system of equations
\eqref{conserv_osc_pert} and \eqref{phi_pert},
describing the infinitesimal radial oscillations of the configuration.
For a polytropic equation of state,
the investigation of this system of equations
is presented in Appendix \ref{appen_osc_poly}.
As in the case of relativistic stars without wormhole
discussed in Ref.~\cite{Chandrasekhar:1964zz},
solving the equations needs a special approach
because of the presence of singular points in the
equations \eqref{conserv_osc_pert_poly} and \eqref{phi_pert_poly}.
An example of a solution is given
in Appendix \ref{appen_osc_poly} (see Fig.~\ref{omega_n_1_5}).

Another important question arising
when considering models of compact objects
is the stability of such configurations.
The question of stability of wormholes with massless ghost fields
or phantom matter was, for instance, investigated in
\cite{ArmendarizPicon:2002km,Bronnikov:1999wh,Lobo:2005yv,Gonzalez:2008wd,Gonzalez:2008xk} (see also references therein).
As discussed in \cite{Gonzalez:2008wd,Gonzalez:2008xk},
the main difficulty of such studies is related to the behaviour of
the perturbations near the throat.
For the ``star plus wormhole" system
in Appendix \ref{appen_stab} a qualitative analysis
of the possibility of the existence of stable solutions
outside the fluid was carried out.
This analysis has shown that the use of a massless ghost scalar field
as a source of nontrivial topology allows for the existence of
an infinite number of oscillating perturbed modes whose energy,
nevertheless, remains finite.
To reach a definite conclusion concerning the stability
of the system under consideration
with respect to radial perturbations of the scalar field,
however, the full system must be analyzed.

The investigations presented in the paper
can be considered as preliminary ones.
They show the principle possibility of the
existence of compact astrophysical objects of the type ``star plus wormhole''.
Further investigations can be directed towards:
(i) a more detailed analysis of such type of configurations
with different parameters of the polytropic equation of state;
(ii) the use of a more realistic equations of state, for example,
for the description of the structure of neutron stars when it
is important to take into account relativistic effects;
(iii) the estimation of the energy associated with
the presence of the oscillations of stellar objects,
and possible applications of such estimates
for a description of various astrophysical events in the Universe
(for example, the gamma-ray bursts);
(iv) the performance of a comparative analysis of the obtained models
with the known models of ordinary stars with trivial topology
for the purpose of identifying the differences,
which can lead to a number of observational consequences.

\section*{Acknowledgements}
V.D. and V.F. are grateful to the Research Group Linkage
Programme of the Alexander von Humboldt Foundation for the
support of this research. They also would like to thank the
Carl von Ossietzky University of Oldenburg for hospitality
while this work was carried out.
B.K. gratefully acknowledges support by the DFG.

\appendix
\numberwithin{equation}{section}

\boldmath
\section{Transformation to a new variable $\theta$}
\unboldmath
\label{appen_transf}

To perform an analysis of the system ``star plus wormhole''
which is described by the Lagrangian \eqref{lagran_wh_star}
and a polytropic equation of state \eqref{eqs_fluid},
it is more convenient to introduce the new variable $\theta$
related to the density $\rho$ at a given point
and the core density $\rho_c$ by
$$
\rho=\rho_c \theta^n,
$$
where $n$ is a constant related to $\gamma$,
Eq.~\eqref{eqs_fluid}, as $n=1/(\gamma-1)$.
Such variable allows to perform a comparative analysis
of the model both for relativistic and non-relativistic polytropic stars.
Then, according to Ref.~\cite{Tooper:1964},
we choose the static metric in the form
\begin{equation}
\label{metric_sphera}
ds^2=e^{\nu(r)}dt^2-e^{\lambda(r)}dr^2
     -r^2\left(d\theta^2+\sin^2\theta d\phi^2\right),
\end{equation}
where the metric functions $\nu, \lambda$ depend only
on the radial coordinate $r$.
The corresponding gravitational equations will then be
\begin{eqnarray}
\label{Einstein-00_app}
&&G_0^0=-e^{-\lambda}\left(\frac{1}{r^2}-\frac{\lambda^\prime}{r}\right)+\frac{1}{r^2}
=8\pi G T_0^0=
8\pi G\left[\rho-\frac{1}{2}e^{-\lambda}\phi^{\prime 2}\right],
 \\
\label{Einstein-11_app}
&&G_1^1=-e^{-\lambda}\left(\frac{1}{r^2}+\frac{\nu^\prime}{r}\right)+\frac{1}{r^2}
=8\pi G T_1^1=
8\pi G\left[-p+\frac{1}{2}e^{-\lambda}\phi^{\prime 2}\right],
\end{eqnarray}
where the energy-momentum tensor from Eq.~\eqref{emt_wh_star}
has been used for obtaining the components $T_0^0$ and $T_1^1$.
The expression for $\phi^{\prime 2}$ can be obtained
by integrating the equation on the massless scalar field
$$
\frac{1}{\sqrt{-g}}\frac{\partial}{\partial x^i}\left[\sqrt{-g}g^{ik}\frac{\partial \phi}{\partial x^k}\right]=0
$$
in the following form
\begin{equation}
\label{first_int}
\phi^{\prime 2}=\frac{D^2}{r^4}e^{\lambda-\nu},
\end{equation}
where $D$ is an integration constant.
Then the system of equations
 \eqref{Einstein-00_app}-\eqref{Einstein-11_app} takes the form:
\begin{eqnarray}
\label{Einstein-00_app_n}
&&-e^{-\lambda}\left(\frac{1}{r^2}-\frac{\lambda^\prime}{r}\right)+\frac{1}{r^2}
=
8\pi G\left[\rho-\frac{1}{2}\frac{D^2}{r^4}e^{-\nu}\right],
 \\
\label{Einstein-11_app_n}
&&-e^{-\lambda}\left(\frac{1}{r^2}+\frac{\nu^\prime}{r}\right)+\frac{1}{r^2}
=
8\pi G\left[-p+\frac{1}{2}\frac{D^2}{r^4}e^{-\nu}\right].
\end{eqnarray}
One more equation follows from the conservation law
 $T^k_{1; k}=0$:
\begin{equation}
\label{conserv_app}
\frac{dp}{dr}=-\frac{1}{2}(\rho+p)\frac{d\nu}{d r}.
\end{equation}

Following Ref.~\cite{Tooper:1964}, we now define a new function $u$ by
\begin{equation}
\label{u_app}
u(r)=\frac{r}{2 G M}\left(1-e^{-\lambda}\right) \rightarrow e^{-\lambda}=1-\frac{2 G M u}{r},
\end{equation}
where $M$ is the mass of the configuration within
the range $r_0 \leq r \leq r_b$,
where $r_b$ is the boundary of the fluid where $\theta=0$.
Using this function, Eq.~\eqref{Einstein-00_app_n} becomes
\begin{equation}
\label{00_via_u}
M\frac{d u}{d r}=4\pi r^2\left(\rho-\frac{1}{2}\frac{D^2}{r^4}e^{-\nu}\right).
\end{equation}
One can see from this equation that $M u(r)=M(r)$
can be interpreted as the total mass of the configuration,
including contributions both from the fluid and the scalar field,
within a sphere of coordinate radius $r$.
For the spherically symmetric case without a wormhole
(what, in our case, corresponds to absence of the scalar field),
one has to put $u(0)=0$ to avoid
a singularity for the mass at the origin \cite{Tooper:1964}.
This corresponds to the fact that the mass at the origin is equal to zero.
In the model with the wormhole,
there exists a minimal value of $r=r_0$,
and correspondingly there exists some minimal mass.
Therefore we have $u(r_0)\neq 0$ (see Eq.~\eqref{bound}).

Next, using \eqref{eqs_fluid_theta}, we have from \eqref{conserv_app}
\begin{equation}
\label{conserv_app_2}
2\sigma (n+1) d\theta+(1+\sigma \theta)d\nu=0,
\end{equation}
with
\begin{equation}
\label{sigma_expr}
\sigma = k \rho_c^{1/n}=\frac{p_c}{\rho_c},
\end{equation}
where $p_c$ is the pressure at the core of the configuration.
Eq.~\eqref{conserv_app_2} may be integrated
to obtain $e^{\nu}$ in terms of $\theta$:
\begin{equation}
\label{nu_app}
	e^{\nu}=e^{\nu_c}\left(\frac{1+\sigma}{1+\sigma \theta}\right)^{2(n+1)},
\end{equation}
where $e^{\nu_c}$ is the value of $e^{\nu}$ at the core of the configuration
where $\theta=1$.
The integration constant $\nu_c$, corresponding to the value of $\nu$
at the  throat,
is determined by requiring $e^{\nu}$ to be equal to unity at infinity
that corresponds to an asymptotically flat space-time.

Next, let us rewrite Eq.~\eqref{Einstein-11_app_n} in the following form:
by using \eqref{00_via_u} one can exclude from
\eqref{Einstein-11_app_n} the term
$$
\frac{1}{2}\frac{D^2}{r^4}e^{-\nu}=\rho-\frac{1}{4\pi r^2} M\frac{d u}{d r},
$$
and express from \eqref{conserv_app_2} the derivative
$$
\nu^\prime=-\frac{2\sigma (n+1)}{1+\sigma \theta} \frac{d\theta}{d r}.
$$
Substituting the two last expressions into
Eq.~\eqref{Einstein-11_app_n},
and using the definition for $e^{-\lambda}$ from \eqref{u_app}, we find
\begin{equation}
\label{Einstein-11_app_n2}
-\left(1-\frac{2G M u}{r}\right)\left[\frac{1}{r^2}-\frac{1}{r}\frac{2\sigma (n+1)}{1+\sigma \theta}\frac{d\theta}{d r}\right]+\frac{1}{r^2}=
8\pi G \left[\rho_c \theta^n(1-\sigma \theta)-\frac{1}{4\pi r^2}M\frac{du}{dr}\right].
\end{equation}
Now we make a change of variables
which puts Eqs.~\eqref{00_via_u} and \eqref{Einstein-11_app_n2}
in dimensionless form, defining
\begin{equation}
\label{dimless_xi_v}
\xi=A r, \quad v(\xi)=\frac{A^3 M}{4\pi \rho_c} u(r), \quad \text{where} \quad A=\left[\frac{4\pi G \rho_c}{(n+1)k \rho_c^{1/n}}\right]^{1/2},
\end{equation}
with $A$ having the dimension of inverse length.
(Here we use the dimensionless variable $v(\xi)$,
according to the notations of Ref.~\cite{Tooper:1964}.
In section \ref{rad_oscill} we have already used the letter $v$
for the designation of the three-velocity.
But this should not lead to confusion.)
Using these new variables $v(\xi)$ and $\xi$,
Eqs.~\eqref{00_via_u} and \eqref{Einstein-11_app_n2} take the form
\begin{eqnarray}
\label{eq_theta_app}
&&\xi^2\frac{1-2\sigma(n+1)v/\xi}{1+\sigma\theta}\frac{d\theta}{d\xi}=
\xi^3\left[\theta^n\left(1-\sigma\theta\right)-\frac{1}{\xi^2}\frac{d v}{d\xi}\right]-v
,\\
\label{eq_v_app}
&&\frac{d v}{d\xi}=\xi^2\left[\theta^n-\frac{1}{2}\frac{\bar{D}^2}{\xi^4}e^{-\nu_c}\left(\frac{1+\sigma}{1+\sigma\theta}\right)^{-2(n+1)}\right],
\end{eqnarray}
with a new dimensionless integration constant
$$
\bar{D}=\frac{4\pi G D}{(n+1)\sigma} \sqrt{\rho_c}\,.
$$
In the absence of the scalar field, i.e.~when $D=0$,
this system reduces to the known system of equations
for the relativistic polytropic gas of Ref.~\cite{Tooper:1964}.
In the non-relativistic limit, i.e.~for $\sigma \to 0$
(corresponding to $p \ll \rho$),
the system of equations \eqref{eq_theta_app}-\eqref{eq_v_app}
without the scalar field reduces to the Lane-Emden equation
$$
\frac{1}{\xi^2}\frac{d}{d\xi}\left[\xi^2\frac{d\theta}{d\xi}\right]=-\theta^n.
$$


\section{Oscillations of the system with a polytropic equation of state}
\label{appen_osc_poly}

As note above, oscillations for various equations of state
of matter creating compact configurations
have been considered extensively earlier.
In particular, for the polytropic equation of state \eqref{eqs_fluid_theta}
used in Ref.~\cite{Chandrasekhar:1964zz}
for the investigation of the dynamical instability of a gas sphere,
the corresponding perturbed component of the pressure $p_p$ is given by
\begin{equation}
\label{press_pert}
p_p=k \rho_c^{1+1/n}(n+1)\theta_0^n \theta_p\,,
\end{equation}
hence
\begin{equation}
\label{press_der_pert}
\frac{\partial p_p}{\partial r}=
k \rho_c^{1+1/n}(n+1)\theta_0^n\left[\theta_p^\prime+n\frac{\theta_0^\prime}{\theta_0}\theta_p\right].
\end{equation}
Taking into account that
\begin{equation}
\label{dens_pert}
\rho_p=n \rho_c \theta_0^{n-1} \theta_p\,,
\end{equation}
we have
\begin{equation}
\label{sum_press_dens_pert}
\rho_p+p_p=\rho_c \theta_0^n\left[\frac{n}{\theta_0}+\sigma(n+1)\right]\theta_p\,.
\end{equation}
Next,
$$
\rho_0+p_0=\rho_c \theta_0^n(1+\sigma \theta_0) \quad \Rightarrow \quad
\rho_0^\prime+p_0^\prime=\rho_c \theta_0^n \theta_0^\prime \left[\frac{n}{\theta_0}+\sigma(n+1)\right].
$$
Using the obtained expressions, equations
\eqref{lambda_pert}-\eqref{nu_prime_pert} can be rewritten as
\begin{eqnarray}
\label{lambda_pert_app}
&&\lambda_p=-8\pi G e^{\lambda_0} r \left[\rho_c \theta_0^n(1+\sigma \theta_0)\zeta+e^{-\lambda_0}\phi_0^\prime \phi_p\right]
,\\
\label{E_pert_app}
&&\theta_p=\frac{1}{n\rho_c \theta_0^{n-1}}\Big\{-\frac{1}{2}e^{-\lambda_0}\phi_0^{\prime 2}\lambda_p-
\frac{2}{r}\left[\rho_c \theta_0^n(1+\sigma \theta_0)\zeta+e^{-\lambda_0}\phi_0^\prime \phi_p\right]\nonumber\\
&&-\rho_c \theta_0^n\left[(1+\sigma \theta_0)\zeta^\prime+\theta_0^\prime\left(\frac{n}{\theta_0}+\sigma(n+1)\right)\zeta\right]
+e^{-\lambda_0}\left[\lambda_0^\prime \phi_0^\prime -\phi_0^{\prime\prime}\right]\phi_p\Big\}
,\\
\label{nu_prime_pert_app}
&&\nu_p^\prime=\nu_0^\prime \lambda_p+\frac{\lambda_p}{r}-8\pi G r e^{\lambda_0}
\left[-k \rho_c^{1+1/n}(n+1)\theta_0^n \theta_p+e^{-\lambda_0}\phi_0^\prime\left(\phi_p^\prime-\frac{1}{2}\phi_0^\prime \lambda_p\right)\right].
\end{eqnarray}
These equations can be rewritten through the dimensionless variables
from \eqref{dimless_xi_v} and new dimensionless variables $\varphi$
for the scalar field and $\psi$ for the displacement
$$
\phi=\frac{\sqrt{\rho_c}}{A}\,\varphi, \quad \zeta=\frac{\psi}{A}.
$$
Taking into account that
(see Eqs.~\eqref{u_app} and \eqref{dimless_xi_v})
$$
e^{-\lambda_0}=1-2\sigma(n+1)\frac{v_0}{\xi} \quad \Rightarrow \quad
\lambda_0^\prime=\frac{2\sigma(n+1)(v_0^\prime-v_0/\xi)}{\xi-2\sigma(n+1)v_0},
$$
and the expression for $\nu_0^\prime$ from \eqref{nu_app}
$$
e^{\nu_0}=e^{\nu_c}\left[\frac{1+\sigma}{1+\sigma \theta_0}\right]^{2(n+1)} \quad \Rightarrow \quad
\nu_0^\prime=-\frac{2\sigma(n+1)}{1+\sigma \theta_0}\theta_0^\prime,
$$
we find from \eqref{lambda_pert_app}-\eqref{nu_prime_pert_app}:
\begin{eqnarray}
\label{lambda_pert_app_dml}
&&\lambda_p=-2\sigma(n+1) e^{\lambda_0} \xi \left[ \theta_0^n(1+\sigma \theta_0)\psi+e^{-\lambda_0}\varphi_0^\prime \varphi_p\right]
,\\
\label{E_pert_app_dml}
&&\theta_p=\frac{1}{n \theta_0^{n-1}}\Big\{-\frac{1}{2}e^{-\lambda_0}\varphi_0^{\prime 2}\lambda_p-
\frac{2}{\xi}\left[ \theta_0^n(1+\sigma \theta_0)\psi+e^{-\lambda_0}\varphi_0^\prime \varphi_p\right]\nonumber\\
&&- \theta_0^n\left[(1+\sigma \theta_0)\psi^\prime+\theta_0^\prime\left(\frac{n}{\theta_0}+\sigma(n+1)\right)\psi\right]
+e^{-\lambda_0}\left[\lambda_0^\prime \varphi_0^\prime -\varphi_0^{\prime\prime}\right]\varphi_p\Big\}
,\\
\label{nu_prime_pert_app_dml}
&&\nu_p^\prime=\nu_0^\prime \lambda_p+\frac{\lambda_p}{\xi}-2\sigma(n+1) \xi e^{\lambda_0}
\left[-\sigma(n+1)\theta_0^n \theta_p+e^{-\lambda_0}\varphi_0^\prime\left(\varphi_p^\prime-\frac{1}{2}\varphi_0^\prime \lambda_p\right)\right],
\end{eqnarray}
where ``prime'' denotes now differentiation with respect
to the dimensionless variable $\xi$.

\begin{figure}[t]
\centering
  \includegraphics[height=6cm]{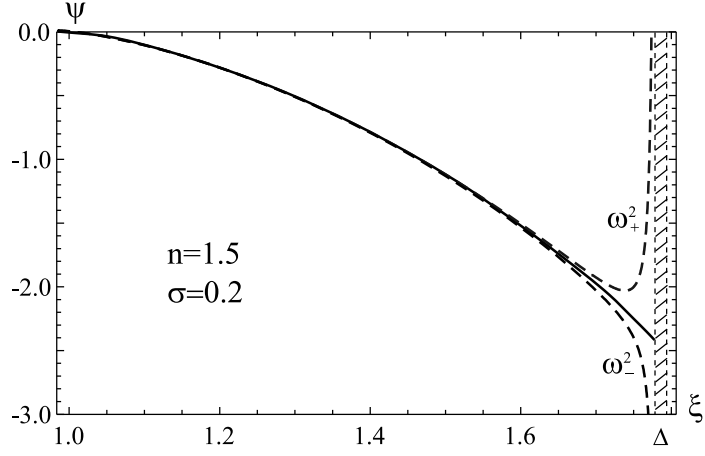}
\caption{The dimensionless displacement $\psi$
obtained by solving the system of
equations \eqref{conserv_osc_pert_poly} and \eqref{phi_pert_poly}
for the case $n=1.5$.
For these curves, the values of $\omega^2$ are the following:
for the top dashed line $\omega_{+}^2=0.177$,
for the bottom dashed line $\omega_{-}^2=0.175$.
The solid line corresponds to $\omega^2=0.1753$ and $\Delta \approx 0.001$.
The boundary of the fluid is located at $\xi_b \approx 1.79$.
}
    \label{omega_n_1_5}
\end{figure}

Using the obtained dimensionless expressions,
Eq.~\eqref{conserv_osc_pert} takes the form
(here a new dimensionless frequency $\bar{\omega}$ is introduced,
$\bar{\omega}=\omega/A$;
we further drop the bar for simplicity)
\begin{align}
\label{conserv_osc_pert_poly}
\begin{split}
	&
	\omega^2 e^{\lambda_0-\nu_0}\left[\theta_0^n(1+
	\sigma \theta_0)\psi+e^{-\lambda_0}\varphi_0^\prime \varphi_p\right]-
	\sigma(n+1)\theta_0^n\left[\theta_p^\prime+
	n\frac{\theta_0^\prime}{\theta_0}\theta_p\right]
\\
&
	+e^{-\lambda_0}\left\{ \left(\varphi_0^{\prime\prime}-
	\lambda_0^\prime \varphi_0^\prime\right)\left(
	\varphi_p^\prime-\frac{1}{2}\varphi_0^\prime \lambda_p\right)+
	\varphi_0^\prime\left[\varphi_p^{\prime\prime}-
	\frac{1}{2}\left(\varphi_0^{\prime\prime}\lambda_p+
	\varphi_0^{\prime}\lambda_p^\prime\right)\right]\right\}
\\
&
	-\frac{1}{2}\theta_0^n\left[\frac{n}{\theta_0}+
	\sigma(n+1)\right]\nu_0^\prime \theta_p+
	e^{-\lambda_0}\left[\varphi_0^\prime \nu_0^\prime \varphi_p^\prime+
	\frac{1}{2}\varphi_0^{\prime 2}\left(\nu_p^\prime-\nu_0^\prime \lambda_p\right)\right]
\\
&
+\frac{4}{\xi}e^{-\lambda_0}\varphi_0^\prime \left(\varphi_p^\prime-\frac{1}{2}\varphi_0^\prime \lambda_p\right)-
\frac{1}{2}\theta_0^n(1+\sigma \theta_0)\nu_p^\prime=0.
\end{split}
\end{align}
The perturbation equation for the scalar field
 \eqref{phi_pert} is then given by
\begin{equation}
\label{phi_pert_poly}
\varphi_p^{\prime\prime}+\left[\frac{2}{\xi}+\frac{1}{2}\left(\nu_0^\prime-\lambda_0^\prime\right)\right]\varphi_p^\prime+
\frac{1}{2}\left(\nu_p^\prime-\lambda_p^\prime\right)\varphi_0^\prime+\omega^2 e^{\lambda_0-\nu_0}\varphi_p=0.
\end{equation}

So we have two linear second-order ordinary differential equations
\eqref{conserv_osc_pert_poly} and \eqref{phi_pert_poly}
for the dimensionless perturbations of the scalar field $\varphi_p$
and the displacement $\psi$.

As an example, let us demonstrate the solution of this system
for the case of $n=1.5$ whose static solutions have been found
in section \ref{stat_sol} (see Fig.~\ref{n_1_5}).
The numerical calculations show that for this case
the solution of the system of
equations \eqref{conserv_osc_pert_poly} and \eqref{phi_pert_poly}
has a singularity at the boundary of the fluid when $\theta \to 0$.
For some value of the frequency $\omega_{+}$
the solution for the displacement $\psi$ tends to $+\infty$,
and for some value $\omega_{-}$ to $-\infty$
(see Fig.~\ref{omega_n_1_5}).
One expects that there exists some value of $\omega$ in the range between
$\omega_{+}$ and $\omega_{-}$ at which the solution becomes regular.
In order to find such solutions, it is possible to use one
of the methods described in Ref.~\cite{Bardeen:1966}.
One such method consists in the following procedure:
determining the eigenfunctions accurately everywhere,
one can integrate simultaneously from the  throat
and from the boundary of the fluid,
and require matching of the logarithmic derivative of $\psi$
at an interior point.
On the other hand, it is possible to use the shooting method
when one can try to find
(using a method of step-by-step approximation)
an eigenvalue of $\omega$ at which the eigenfunctions are regular.
Using such an approach, and choosing $\omega$, one can try to achieve
that the solution will be approaching closer and closer to
the boundary of the fluid $\xi_b$
until it will not reach some critical value $\xi_{crit}$
at which the solution diverges.
The width $\Delta=\left(\xi_b-\xi_{crit}\right)$ is determined by:
(i) the structure of equations \eqref{conserv_osc_pert_poly}
and \eqref{phi_pert_poly};
(ii) the accuracy of the numerical method used;
(iii) the machine precision.
Using the NDSolve routine from {\it Mathematica},
we have reduced $\Delta$ to $\approx 0.001$ at $\omega^2 \approx 0.1753$.
The solution for this value of $\omega^2$ is shown
in Fig.~\ref{omega_n_1_5} by the solid line.

\section{Stability analysis}
\label{appen_stab}

In this section we study the dynamical stability of the above static
solutions under linear perturbations.
Due to considerable technical challenges 
in performing the stability analysis for the system as a whole,
we just present a linear stability analysis
of the external solution outside the fluid.
Restricting ourselves to this region,
makes the calculations much simpler,
while it is still possible to demonstrate some general properties
concerning the stability of the system under consideration.

Outside the fluid, the static solutions are given by the system of equations
\eqref{eq_v_ext}-\eqref{eq_phi_ext}.
The corresponding time-dependent equations outside the fluid
can be obtained from equations \eqref{Einstein-00osc}-\eqref{emt-10}
by putting $\rho=p=0$.
The time-dependent equation for the scalar field can be written as follows:
\begin{equation}
\label{sf_ext_stab}
e^{-\nu}\left[\ddot{\phi}-\frac{1}{2}\left(\dot{\nu}-\dot\lambda\right)\dot{\phi}\right]-
e^{-\lambda}\left\{\phi^{\prime\prime}+\left[\frac{2}{r}+\frac{1}{2}\left(\nu^\prime-\lambda^\prime\right)\right]\phi^\prime\right\}=0.
\end{equation}
We perturb the solutions of this system
by expanding the metric functions and the scalar field
function to first order as follows:
\begin{equation}
\label{pertubs}
\lambda=\lambda_0(r)+\lambda_1(r)\cos{\omega t}, \quad
\nu=\nu_0(r)+\nu_1(r)\cos{\omega t}, \quad
\phi=\phi_0(r)+\phi_1(r)\frac{\cos{\omega t}}{r}.
\end{equation}
The index 0 indicates the static background solutions,
and the index 1 refers to perturbations.
Substituting these expressions into the
Einstein equations \eqref{Einstein-00osc}-\eqref{Einstein-10osc},
one can find the following expressions for the metric perturbations:
\begin{equation}
\label{lambda_pert_2}
\lambda_1=-8\pi G \phi_0^\prime \phi_1
\end{equation}
and
\begin{equation}
\label{nu_pert}
\nu_1^\prime=\lambda_1^\prime-\frac{8\pi G \phi_0^\prime}{r}\left[2+r\left(\nu_0^\prime-\lambda_0^\prime\right)\right]\phi_1.
\end{equation}

Then, substituting the perturbations \eqref{pertubs}
and expressions \eqref{lambda_pert_2} and \eqref{nu_pert}
into Eq.~\eqref{sf_ext_stab},
one can obtain the following equation for $\phi_1$
rewritten in the dimensionless variables used in the previous sections:
\begin{equation}
\label{phi_pert_poly_stab_n}
\varphi_1^{\prime\prime}+\frac{1}{2}\left(\nu_0^\prime-\lambda_0^\prime\right)\varphi_1^\prime-V_0(\xi)\varphi_1+
\bar{\omega}^2 e^{\lambda_0-\nu_0}\varphi_1=0,
\end{equation}
with the potential given by
\begin{equation}
\label{poten_stab}
V_0(\xi)=\frac{1}{2}\frac{\nu_0^\prime-\lambda_0^\prime}{\xi}+\sigma(n+1)
\varphi_0^{\prime 2}\left[2+\xi\left(\nu_0^\prime-\lambda_0^\prime\right)\right],
\end{equation}
and $\bar{\omega}=\omega/A$.
Introducing the new independent variable $\eta$
\begin{equation}
\label{new_var_stab}
\frac{d\eta}{d\xi}=\exp{\left[\frac{1}{2} \left(\lambda_0-\nu_0\right)\right]},
\end{equation}
one can rewrite Eq.~\eqref{phi_pert_poly_stab_n} in a Schr\"{o}dinger-like form
\begin{equation}
\label{phi_pert_poly_stab_n2}
-\frac{d^2\varphi_1}{d\eta^2}+V[\xi(\eta)]\varphi_1=\bar{\omega}^2\varphi_1,
\end{equation}
where
\begin{equation}
\label{poten_stab_2}
V[\xi(\eta)]=e^{\nu_0-\lambda_0} V_0(\xi).
\end{equation}
Now the consideration of the stability of the system
reduces to a study of Eq.~\eqref{phi_pert_poly_stab_n2}.
Then if the energy-like eigenvalues $\bar{\omega}^2$ are positive,
the solution is stable w.r.t.~the perturbations addressed,
due to the regularity of the perturbations
(see the expressions for the perturbations \eqref{pertubs}).
Another important point, providing regularity,
is the localization of the eigenfunction $\varphi_1$
in a bounded region of space.
The qualitative behavior of the solutions of Eq.~\eqref{phi_pert_poly_stab_n2}
can be estimated from the form of the potential \eqref{poten_stab_2}.
As an example, in Fig.~\ref{poten_stab-fig}
the form of this potential for the case considered
in Appendix \ref{appen_osc_poly} with $n=1.5, \sigma=0.2$
(see Fig.~\ref{omega_n_1_5}) is presented.
One can see from Fig.~\ref{poten_stab-fig} that,
using the above-mentioned parameters,
the potential \eqref{poten_stab_2} looks like a wall
near the boundary of the fluid at $\xi=\xi_b$
(the variable $\eta$ is chosen in such a way as to provide $\eta_b=\xi_b$
at the boundary of the fluid).
In quantum mechanical language,
this will correspond to motion of a particle
in the field of the potential \eqref{poten_stab_2}
with an oscillating eigenfunction $\varphi_1$.
The absence of a well in the potential \eqref{poten_stab_2}
inhibits the existence of a discrete spectrum of energy levels
of the system \eqref{phi_pert_poly_stab_n2}-\eqref{poten_stab_2}.

\begin{figure}[t]
\centering
  \includegraphics[height=5cm]{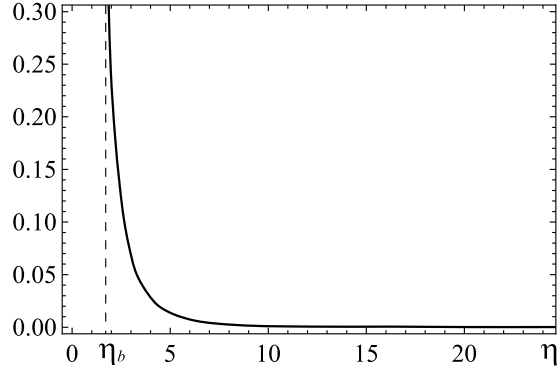}
\caption{The form of the potential \eqref{poten_stab_2}
for the case  of $n=1.5, \sigma=0.2$.
The potential is always positive for these parameters.
The left boundary is situated at the edge of the fluid
at $\eta_b=\xi_b \approx 1.79$.
Asymptotically, at $\eta \to \infty$, the potential goes to zero.
}
    \label{poten_stab-fig}
\end{figure}

Let us now estimate energy associated with these oscillations.
The energy density of the scalar field is
\begin{equation}
\label{energ_dens_phi}
\varepsilon_{\phi}=\left[T_0^0\right]_{\phi}=-\frac{1}{2}\left(e^{-\nu}\dot\phi^2+e^{-\lambda}\phi^{\prime^2}\right).
\end{equation}
Then using the expressions for perturbations from \eqref{pertubs}, one can obtain the perturbed part of \eqref{energ_dens_phi}
$\left[T_0^0\right]_{p}$
in the form
$$
\left[T_0^0\right]_{p}=-e^{-\lambda_0}\phi_0^{\prime}\left[\frac{1}{r}\left(\phi_1^\prime-\frac{\phi_1}{r}\right)-\phi_0^\prime
\frac{\lambda_1}{2}\right]\cos {\omega t},
$$
or in dimensionless variables
\begin{equation}
\label{energ_dens_phi_dim}
\left[T_0^0\right]_{p}=-\rho_c e^{-\lambda_0}\varphi_0^{\prime}\left[\frac{\mathfrak{D}}{\xi}\left(\phi_1^\prime-\frac{\phi_1}{\xi}\right)-\varphi_0^\prime
\frac{\lambda_1}{2}\right]\cos {\bar{\omega} \tau},
\end{equation}
where $\tau=A t$ is the dimensionless time variable, and a new dimensionless constant
$$
\mathfrak{D}=\frac{A^2}{\sqrt{\rho_c}}=\frac{4\pi G \sqrt{\rho_c}}{\sigma(n+1)}
$$
is introduced. Then the proper energy of perturbations is defined by the integral
$$
E=\int_{r_b}^{\infty} \sqrt{-g^{(3)}}\left[T_0^0\right]_{p} d^3 x,
$$
or in dimensionless variables
\begin{equation}
\label{energ_phi_dim}
\tilde{E}=4\pi \cos^2 \bar{\omega} \tau \int_{\xi_b}^{\infty}
\xi^2 \varphi_0^\prime e^{-\lambda_0/2}\left[\varphi_0^\prime\frac{\lambda_1}{2}-\frac{\mathfrak{D}}{\xi}
\left(\phi_1^\prime-\frac{\phi_1}{\xi}\right)\right]d\xi,
\end{equation}
where $\tilde{E}=E/E_c$ is the dimensionless energy measured in units of some characteristic energy $E_c=\rho_c/A^3$.
To show that the oscillations have finite energy,
we can perform an analytical estimation of the asymptotic value of this integral.
To do this, let us find  asymptotic expressions for the functions $\varphi_0^\prime,  \lambda_0, \lambda_1$ and $\phi_1$
appearing in the integral.
It follows from equation \eqref{eq_v_ext} that asymptotically
\begin{equation}
\label{v_0_asymp}
\frac{d v_0}{d\xi} \approx -\frac{1}{2}\frac{\bar{D}^2}{\xi^2} \quad \Rightarrow \quad
v_0 \approx \frac{\bar{D}^2}{2\xi}.
\end{equation}
Then from \eqref{lambda_poly} we have
$$
e^{\lambda_0}\approx 1+\sigma(n+1)\frac{\bar{D}^2}{\xi^2}.
$$
Next, equation  \eqref{eq_phi_ext} gives
$$
\varphi_0^\prime \approx \frac{\bar{D}}{\xi^2}.
$$
The asymptotic expression for  $\phi_1$ can be found as follows:
one can see from the  form of the potential \eqref{poten_stab_2} that it
approaches zero asymptotically. Correspondingly,  equation
 \eqref{phi_pert_poly_stab_n2}  describes motion of a free particle, and its solution
can be presented in the form
$$
\phi_1 \approx \alpha \cos{(\bar{\omega}\xi-\beta)},
$$
where $\alpha, \beta$ are some arbitrary constants. Then the expression for $\lambda_1$ from \eqref{lambda_pert_2} takes the form
$$
\lambda_1 \approx -8\pi G \sqrt{\rho_c} \bar{D} \frac{ \alpha \cos{(\bar{\omega}\xi-\beta)}}{\xi^2}.
$$
Using all these asymptotic expressions, we have the following asymptotic form of the integral from \eqref{energ_phi_dim}:
\begin{equation}
\label{energy_asymp}
\tilde{E}\approx 4\pi
\alpha\bar{D}\mathfrak{D} \bar{\omega} \cos^2 \bar{\omega} \tau  \int_{\xi_a}^{\infty} \frac{\sin{(\bar{\omega}\xi-\beta)}}{\xi}d\xi,
\end{equation}
where the initial  integration point
$\xi_a$ is chosen at that place where the potential
  \eqref{poten_stab_2} is close to zero.
Since this integral is converging, the proper energy of the oscillations remains finite.
Thus, taking into account that the value  $\bar{\omega}^2$ is always positive, one can conclude that the external solution is
stable against these linear perturbations.

\end{document}